\documentclass[amsfonts,amsmath,prl,twocolumn]{revtex4-1}
\usepackage{graphicx}
\usepackage{color}
\usepackage{units}
\usepackage{dsfont}

\newcommand{\tcg}[1]{}

\newcommand{\expec}[1]{\langle#1\rangle}  

\newcommand{\var}[1]{\Delta^2{#1}}

\newcommand{\vop}[1]{\ensuremath{\hat{\boldsymbol{#1}}}}
\newcommand{\absq}[1]{\ensuremath{\bigl|#1\bigr|^2}}

\newcommand{\figref}[1]{Fig.~\ref{fig:#1}}

\begin{document}
\title{Quantum-Dense Metrology}

\author{Sebastian Steinlechner}
\author{J\"oran Bauchrowitz}
\author{Melanie Meinders}
\author{Helge M{\"u}ller-Ebhardt}
\author{Karsten Danzmann}
\author{Roman Schnabel}
\affiliation{Institut f\"ur Gravitationsphysik, Leibniz Universit\"at Hannover and Max-Planck-Institut f\"ur
Gravitationsphysik (Albert-Einstein-Institut), Callinstr. 38, 30167 Hannover, Germany}

\date{\today}

\begin{abstract}
Quantum metrology utilizes entanglement for improving the sensitivity
of measurements \cite{Giovannetti2011,Schnabel2010}.  Up to now the focus has
been on the measurement of just \emph{one} out of two non-commuting
observables. Here we demonstrate a laser interferometer that provides
information about \emph{two} non-commuting observables, with uncertainties
below that of the meter's quantum ground state.  Our experiment is a
proof-of-principle of quantum dense metrology, and uses the additional
information to distinguish between the actual phase signal and a parasitic
signal due to scattered and frequency shifted photons.  Our approach can be
readily applied to improve squeezed-light enhanced gravitational-wave detectors
at \emph{non}-quantum noise limited detection frequencies in terms of a sub
shot-noise veto-channel.
\end{abstract}

\maketitle

Heisenberg's uncertainty principle states that it is generally not possible to
gather precise information about non-commuting observables of a physical
system. Prominent examples are the position and the momentum of a particle or
the amplitude and phase quadratures of an electro-magnetic wave. In this way
the uncertainty of meter systems, e.g. laser light, limits the sensitivity in
metrology, even if the quantum mechanical uncertainty of the measurement object
itself can be neglected. Using a meter system in a nonclassical state it is,
nevertheless, possible to measure one observable with arbitrarily high
precision. If its imprecision is `squeezed' below the zero-point fluctuation of
the meter system the regime of quantum metrology is reached, as demonstrated in
proof-of-principle experiments
\cite{Xiao1987,Grangier1987,Leibfried2004,Cappellaro2005,Vahlbruch2005,Afek2010,Wasilewski2010,Lucke2011}.
Recently, quantum metrology was applied to an operating gravitational wave
detector \cite{Abadie2011}.

For a wide application of quantum metrology a rather general problem exists. In
order to improve classical state-of-the-art measurement sensitivities, the
concept of quantum metrology must be combined with state-of-the-art intense
meter states. Current gravitational wave detectors, be they squeezed-light
enhanced or not, use light fluxes of about $10^{20}$ photons per second
\cite{Abramovici1992}.  Unfortunately, the scattering of just a single photon
from the meter into the signal band per second and Hertz produces a significant
parasitic signal against which quantum-noise squeezing is bootless. The scatter
problem is understood as a \emph{parasitic interference}, where vibrating
scatter surfaces frequency-shift a tiny amount of photons into the detection
band \cite{Vahlbruch2007}. It is a well-known problem in high-precision laser
interferometry \cite{Vinet1996,Vinet1997,Ottaway2012}.  We conjecture that the
limitation of quantum metrology at lower detection frequencies as observed in
\cite{Abadie2011} at least partially originates from parasitic interferences.
In the future, even higher photon fluxes will be used \cite{Hild2010}, and
parasitic interferences will become increasingly severe.

Here we propose \emph{quantum-dense metrology} (QDM) to widen the application
of quantum metrology into the regime where parasitic interferences are a
limiting noise source. We present a proof of principle experiment that
discriminates between the actual science signal and a parasitic interference
with sub-shot-noise measurement precision, exploiting the generally different
phase space orientation of the two. It is shown that QDM provides a
non-classical veto-channel during signal searches and is thus able to improve a
non-classical interferometer beyond what is possible with conventional quantum
metrology. 

\begin{figure}[tp]
    \includegraphics{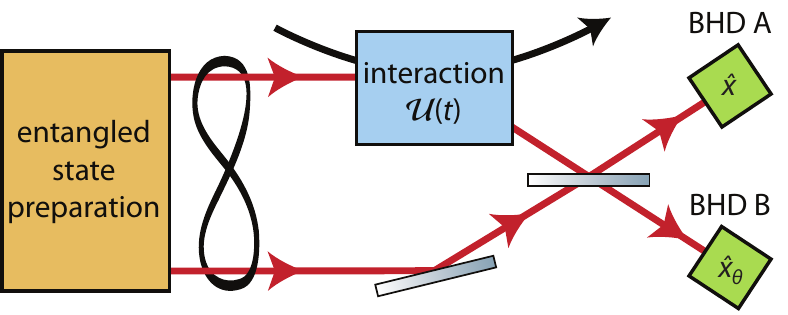}
    \caption{Schematic showing the underlying principle of quantum-dense
    metrology. The meter system consists of a bipartite continuous-variable
    entangled state, of which one part interrogates the system of
    interest in an interaction zone, while the other part is kept as
    a local reference. Both entangled modes are recombined after the interaction, and
    each output is detected via a balanced homodyne detector (BHD).}
    \label{fig:qdm}
\end{figure}

Our readout scheme is based on Einstein-Podolsky-Rosen entanglement
\cite{Einstein1935}, which has been first considered for metrology by D'Ariano
\emph{et al.} \cite{D'Ariano2001}. Following this work we replace the single
meter state by a bipartite, two-mode-squeezed entangled state, as depicted in
\figref{qdm}. One mode of the entangled system serves as the new meter state,
whereas the other mode is kept as an external reference for the measurement
device. Since the difference in position and the sum in momentum commute,
$[\hat x_A - \hat x_B, \hat p_A + \hat p_B] = 0$, it is in principle possible
to exactly measure the distance in phase space between the two modes. Thus we
overcome the limitation that is set by the Heisenberg Uncertainty Relation for
reading out two orthogonal quadratures of a \emph{single} system by performing
all measurements in relation to the reference beam. Such measurements have
previously also been considered for super-dense coding with the purpose of
doubling the capacity of quantum communication channels
\cite{Bennett1992,Braunstein2000}. The required continuous-variable entangled
states were experimentally pioneered by Ou \emph{et al.} \cite{Ou1992}, see
also \cite{Furusawa1998,Silberhorn2001,Bowen2003b} for more recent experiments.
In contrast to all previous proposals, QDM as introduced here uses two-mode
squeezing of \emph{non}-orthogonal quadratures. We show that this opens a way
to optimize the science signal-to-noise ratio.

The measurement problem of reading out two orthogonal quadratures was pioneered
by Arthurs and Kelly \cite{Arthurs1965}. For a light field whose (classical)
signal is given by a displacement projected onto the quadratures $X(t) =
\expec{\hat x (t)}$ and $P(t) = \expec{\hat p (t)}$, the signal-normalized
Heisenberg uncertainty relation reads
\begin{equation}
    \frac{\var{\hat x(t)}}{\absq{X(t)}}\frac{\var{\hat p(t)}}{\absq{P(t)}}
    \geq
    \frac{1}{4\absq{X(t)}\absq{P(t)}}\,,
    \label{eq:heisenberg}
\end{equation}
where the quadrature variances for the ground state are normalized to
$\var{\hat x(t)}=\var{\hat p(t)}=1/2$. However, actually measuring both
quadratures simultaneously (subscript `sim') with e.g.\ an eight-port homodyne
detector \cite{Leonhardt1997} leads to an uncertainty relation that is four times
as large \cite{Arthurs1965},
\begin{equation}
    \frac{\var{\hat x_{\rm sim}(t)}}{\absq{X_{\rm sim}(t)}}
    \frac{\var{\hat p_{\rm sim}(t)}}{\absq{P_{\rm sim}(t)}}
    \geq
    \frac{1}{\absq{X(t)}\absq{P(t)}}\,.
    \label{eq:sim_meas}
\end{equation}

With QDM as presented here, the simultaneous readout is no longer limited by
such an uncertainty relation. Instead, the achievable sensitivity is directly
connected to the squeezing parameters $r_a$, $r_b$ of the initial squeezed
beams. Entangling those beams with relative angle $\theta$ allows for a
simultaneous detection of the quadrature $\hat x$ and the rotated quadrature
$\hat x_\theta = \hat x \cos \theta + \hat p \sin \theta$ with
\begin{equation}
    \frac{\var{\hat x_{\rm sim}^{\rm ent}(t)}}{\absq{X_{\rm sim}^{\rm ent}(t)}}
    \frac{\var{\hat x_{\theta,{\rm sim}}^{\rm ent}(t)}}{\absq{X_{\theta, {\rm
    sim}}^{\rm ent}(t)}}
    \geq
    \frac{e^{-2r_a}e^{-2r_b}}{\absq{X(t)}\absq{X_{\theta}(t)}}\,.
    \label{eq:qdm}
\end{equation}
Setting $\theta = \pi/2$ the substantial improvement compared to the lower
bound in inequality \eqref{eq:sim_meas} becomes obvious. The Heisenberg
uncertainty relation for a conventional readout based on a single-mode meter
system \eqref{eq:heisenberg} is surpassed for two-mode squeezing with $r >
0.3466$. In principle, QDM allows for a readout with arbitrary precision, in
the limit of infinite squeezing. A detailed derivation of the above results can
be found in the supplementary materials.


\begin{figure}[bp]
    \includegraphics{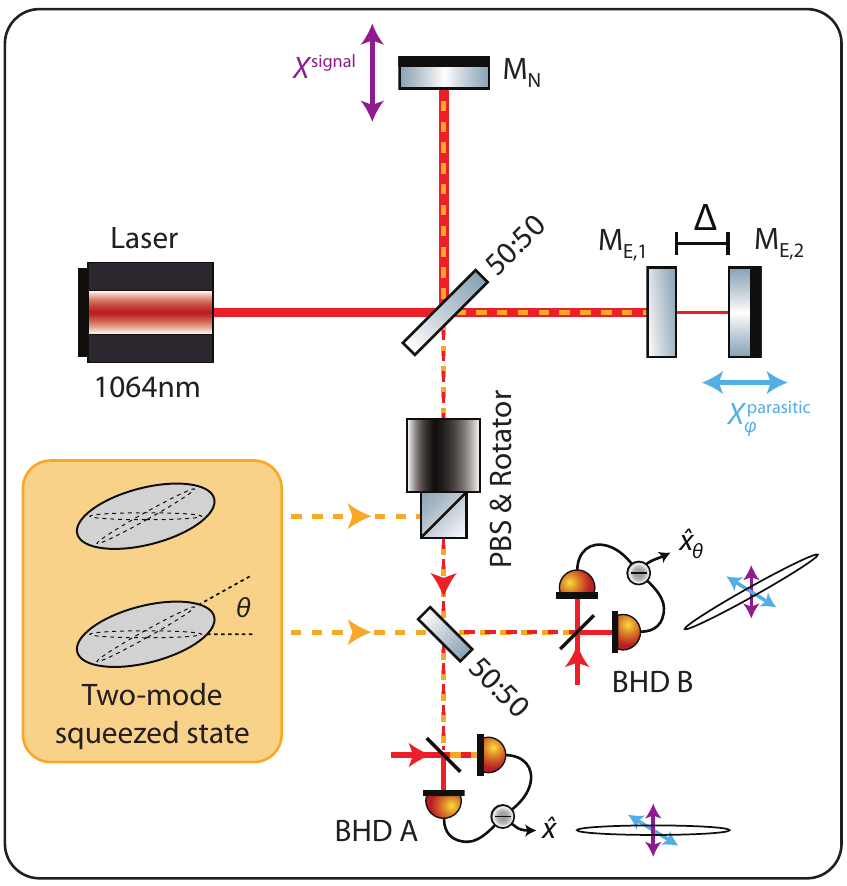}
    \caption{Schematic setup for the experimental demonstration of quantum
    dense metrology. Two signals $X^\text{signal}(t)$ and
    $X^\text{parasitic}_\phi(t)$ are generated in a Michelson interferometer by
    modulating the PZT mounted mirrors $\rm M_N$ and $\rm M_{E,2}$,
    respectively. $X^\text{signal}(t)$ is a pure phase modulation, while
    $X_\phi^\text{parasitic}(t)$ can be rotated into an arbitrary quadrature by
    adjusting $\Delta$, which is the microscopic spacing between the east
    mirrors. A Faraday rotator couples one part of the entangled state into the
    interferometer. The other part is overlapped with the signal leaving the
    interferometer. The two resulting beams are simultaneously detected with
    balanced homodyne detectors, BHD A \& B, measuring quadratures $\hat x$ and
    $\hat x_{\theta \neq 0}$, respectively $\hat p = \hat x_{\theta = \pi/2}$. Revealing
    a parasitic interference requires neither the knowledge of $\phi$ nor
    matching $\theta$ to $\phi$.} 
    \label{fig:setup}
\end{figure}

We proved the principle of quantum-dense metrology and its high potential for
improving state-of-the-art laser interferometers in the following table-top
experiment. In a Michelson-type laser interferometer (\figref{setup}) with arm
lengths of about \unit[7.5]cm we generated two signals in the megahertz regime.
The actual interferometer phase signal was produced by modulating the PZT
mounted north-arm mirror at \unit[5.55]MHz. We intentionally introduced a
parasitic signal at \unit[5.17]MHz by PZT-modulating a small amount of light
that leaked through the east-arm mirror. By adjusting the phase with which the
light was back-reflected into the interferometer, we were able to simulate a
parasitic interferences in any quadrature.  The entangled light was generated
from two squeezed modes following the scheme in \cite{Furusawa1998}.  One mode
of the entangled state was introduced into the interferometer dark port via a
polarizing beam-splitter (PBS) and a Faraday rotator. The output field was
transmitted by the PBS and was overlapped at a $50:50$ beam-splitter with the
other entangled mode. Both beam-splitter outputs were simultaneously detected
via balanced homodyne detection (BHD). A more detailed explanation of the
experimental setup is given in the methods.

\begin{figure}[tbp]
    \includegraphics{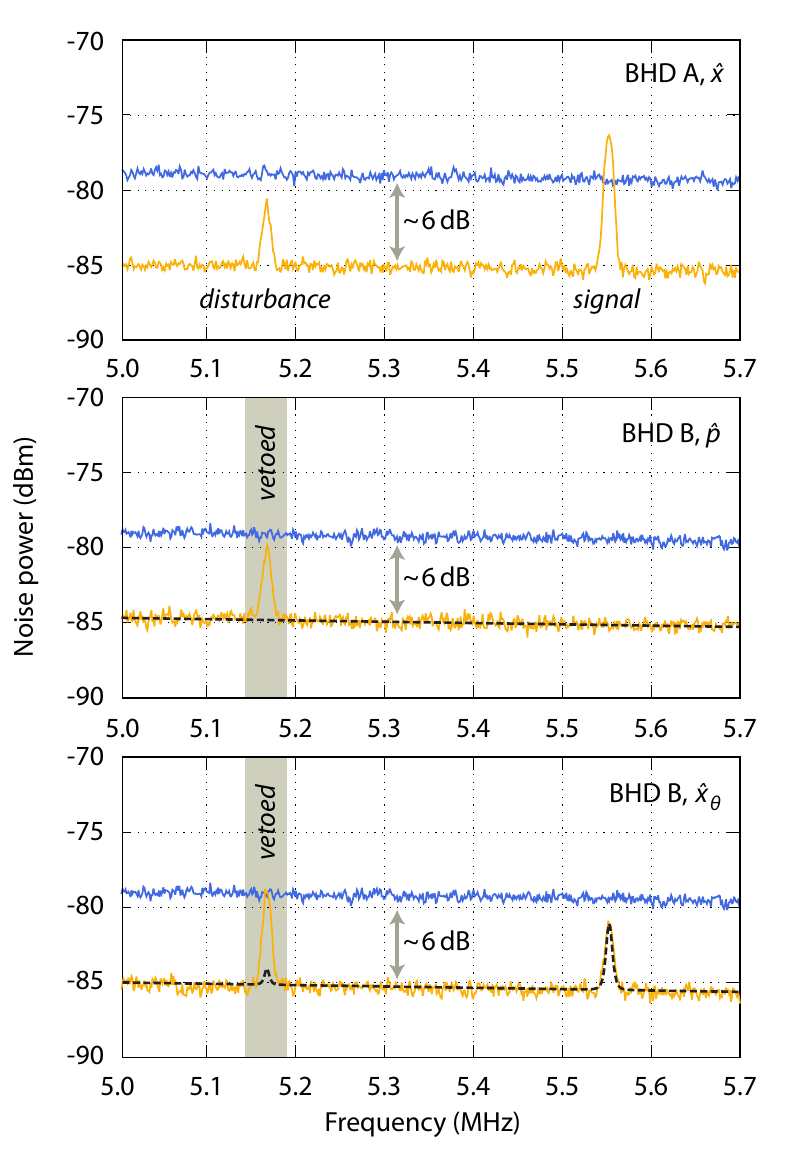}
    \caption{Experimental demonstration of quantum-dense metrology for the
    identification of parasitic interferences. The orange traces show the
    detected signal variances at BHD~A, $\var{\hat x}$ (top panel) and BHD~B,
    $\var{\hat p}$ (middle) and $\var{\hat x_\theta}$ (bottom).  Parasitic
    signals are revealed due to their unexpected scaling in the lower two
    panels.  The calculated scalings of science signals are shown in the dashed
    black curves. In the bottom panel, the angle $\theta$ was tuned so that
    part of the signal was recovered. The blue traces show the vacuum noise
    power of our BHDs, which are slightly sloped due to the decreasing transfer
    function of the homodyne detectors. All traces were recorded with RBW
    \unit[10]kHz, VBW \unit[100]Hz and averaged three times. 
    }
    \label{fig:entifo}
\end{figure}

The results of our experiment are presented in \figref{entifo}.  The first
panel shows the power spectrum of the amplitude quadrature measurement of
BHD~A, which generally provides the highest signal-to-noise-ratio for the
interferometric science signal (here at \unit[5.55]MHz, orange trace).  Due to
the injected entangled light the readout noise is reduced to \unit[6]dB below
the vacuum noise (blue trace).  BHD~A also clearly detects a second (parasitic)
signal at \unit[5.17]MHz. Only by simultaneously looking at $\hat p$, as it is
done with BHD~B (second panel), the parasitic nature of the lower frequency
signal is revealed: while the phase signal at \unit[5.55]MHz vanishes as
expected, the signal at \unit[5.17]MHz does not vanish but actually increases
in size.  This information is sufficient to reveal the parasitic nature of the
lower frequency signal. Also the second BHD shows readout noise roughly
\unit[6]dB below vacuum noise. Simultaneous squeezing in two orthogonal
quadratures is unique to our quantum-dense readout scheme.

In the third panel of \figref{entifo}, we used an improved strategy to reveal
the parasitic signal.  We detuned the angle $\theta$ between the original
squeezing ellipses away from $90^\circ$. This way it is possible to retain at
least part of the science signal in BHD~B, while still having insight into the
orthogonal quadrature. Since it can be exactly calculated how a phase signal
measured at BHD~A is projected into the $\hat x_\theta$ quadrature, any
discrepancy reveals a parasitic signal.  The dashed black lines in the second
and third panels show the projected noise power, assuming that the first panel
contains only phase signals. While the signal at \unit[5.55]MHz perfectly
matches the expectation, the disturbance at \unit[5.17]MHz clearly does not.
The advantage of the measurement in third panel is that, together with the
first panel, the overall signal-to-noise-ratio of the science signal is
improved. Changing $\theta$ allows for a smooth tuning between full signal
coverage ($\theta = 0$), but no information about the conjugate observable; and
maximum information about the disturbances in the full phase space ($\theta =
\pi/2$), but loss of half the science signal power.


\textbf{Conclusions} ---
Interference is the basis for many high precision measurements and parasitic
interferometer signals are a general problem that hampers the usefulness of
nonclassical approaches.  We have introduced and experimentally demonstrated
the concept of quantum-dense metrology. QDM makes use of the fact that the
scientific signal of an interferometer generally appears in a well-determined
quadrature. A parasitic interference, however, appears in an arbitrary
quadrature orientation. QDM can distinguish between scientific and parasitic
signals with unlimited precision beyond the meter's ground state uncertainty,
and we proposed and demonstrated that this can be used to create a veto channel
for parasitic signals. Our approach
uses steady-state entanglement and therefore does not rely on any kind of
conditioning or post-selection, which would result in a loss of measurement
time. For the first time we propose
two-mode squeezing for metrology that is generated with a non-orthogonal
relative squeezing angle. Such entangled states allow the optimization of the
signal-to-noise ratio when QDM is applied.

Beyond what we have demonstrated here, it should be even possible to subtract
parasitic signals from the measurement data without subtracting science
signals. For this, two assumptions have to be made. First, the parasitic
signals have a quasi-stationary phase space orientation, second, the science
signals have a temporal or spectral shape that is different from the parasitic
signal.  Then, quantum tomography at the second balanced homodyne detector (B)
can be used to gather information about the parasitic signal's phase space
orientation and its projected quadrature components. Another idea is keeping
the local oscillator phase fixed and introducing a fitting parameter that
describes by which magnitude the parasitic signal is projected onto the
conventional readout quadrature of the interferometer. Fitting parameters are
already used in data analysis based on matched filtering and signal templates
\cite{Sathyaprakash2009}. In both scenarios, QDM allows for sub shot-noise
measurements even if the apparatus without QDM is limited by parasitic
interferences, i.e.\ is not quantum noise limited.  QDM as proposed here does
not help in the case of pure parasitic \emph{phase} signals, which are caused
by thermally excited fluctuations of mirror surfaces and radiation pressure
forces. Instead, it is a valuable tool against all types of parasitic signals
having a phase space orientation different from the phase quadrature.  Our
scheme can be applied to high-precision laser interferometers such as
gravitational-wave observatories, where it has high potential in identifying
parasitic signals due to photon scattering or hitherto unknown mechanisms. We
thus envision that QDM will widen the application of quantum metrology in
ongoing and future high precision measurements.

\textbf{Acknowledgements} --- We acknowledge discussions with Tobias Eberle, Vitus
H\"andchen and Harald L\"uck. This research has been
financed by the Deutsche Forschungsgemeinschaft (SFB
TR7, project C8), the EU FP-7 project Q-ESSENCE and supported by
the Centre for Quantum Engineering and Space-Time Research (QUEST) and the
IMPRS on Gravitational Wave Astronomy.

\bibliography{EntangledIFO}


\section{Supplementary Material}

\textbf{Entangled-light generation.}
Our continuous-variable entangled light was generated by the source described
in \cite{Steinlechner2011}. Two squeezed vacuum fields generated by type I
parametric down-conversion in PPKTP were overlapped at a $50: 50$ beam
splitter, thereby creating two-mode squeezed light. Both input fields carried a
residual phase modulation from locking the optical parametric amplifiers. At
the detection stage, this modulation was reused to align the homodyne detectors
to the squeezed quadratures. A single sideband modulation was imprinted on one
of the squeezed fields by overlapping it with \unit[80]MHz frequency-shifted
light from an acousto-optical modulator. This sideband was used to lock the
quadrature angle between the input squeezed states. It was also used to
stabilize one mode of the entangled field to the Michelson interferometer by
detecting the beat signal between the sideband and the interferometer input
field behind one end-mirror.

\textbf{Interferometer setup and control.}
The Michelson interferometer had an arm length of about \unit[7.5]cm for the
north arm. The east arm was about \unit[1.5]cm shorter, which allowed us to use
the so-called Schnupp modulation technique \cite{Schnupp1988} for locking the
interferometer to its dark fringe. Both end-mirrors were flat and had a power
reflectivity of $99.98\%$ ($M_N$) and $98\%$ ($M_{E1}$). The north mirror was
PZT mounted to create a phase modulation inside the interferometer. A second
PZT mounted flat mirror $M_{E2}$ with reflectivity of $\approx 20\%$ was placed
a few millimeters behind $M_{E1}$, creating a (weakly coupled) Fabry-P\'erot
cavity. By tuning this cavity, the phase signal created by $M_{E2}$ could be
rotated into an arbitrary quadrature. A DC locking scheme detected the
transmitted light and held the cavity on its operating point. Both PZTs were
driven on a mechanical resonance to be able to create signals in the few
megahertz regime where the detected squeezing was strongest.

\textbf{Readout of orthogonal quadratures.}
Consider a continuous-wave laser beam with central frequency $\omega_0$. The
quantum noise of this beam at the sideband frequencies $\pm \Omega$, measured
with a resolution bandwidth of $\Delta\Omega$, can be described by
time-dependent operators for the amplitude quadrature $\hat a_1(\Omega,
\Delta\Omega, t)$ and the phase quadrature $\hat a_2(\Omega, \Delta\Omega, t)$.
Here we restrict ourselves to a monochromatic signal at a fixed sideband
frequency and therefore drop the explicit frequency dependency in the following
treatment. The quadrature operators satisfy the commutation relation $[\hat
a_1, \hat a_2] = i$ and are normalized such that for a (squeezed) minimum
uncertainty state $\var{\hat a_1} = \var{\hat a_2} = e^{\mp 2r}/2$, where the
minus sign in the exponent belongs to the amplitude quadrature and the plus
sign to the phase quadrature. $r$ is the squeezing parameter, therefore $r=0$
corresponds to a vacuum state, while $r<0$ and $r>0$ correspond to phase and
amplitude squeezed light, respectively.

A measurement adds (classical) amplitude and phase modulations $X(t)$, $P(t)$
to the laser beam. The output field can then be described by the field
quadrature vector
\begin{equation}
    \vop m = \begin{pmatrix}
            \hat x(t) \\
            \hat p(t)
        \end{pmatrix}
        =
        \begin{pmatrix}
            \hat a_1(t) + X(t) \\
            \hat a_2(t) + P(t)
        \end{pmatrix}\,.
\end{equation}
From the commutation relation we can infer the Heisenberg uncertainty relation
for the shot noise, normalized to the signal,
\begin{equation}
    \frac{\var{\hat x(t)}}{\absq{X(t)}}
    \frac{\var{\hat p(t)}}{\absq{P(t)}}
    \geq
    \frac{1}{4\absq{X(t)}\absq{P(t)}} \,.
\end{equation}
This inequality limits the simultaneous measurability of the amplitude and
phase quadrature modulations. A simple approach to actually measure both
quadratures in an Arthurs-Kelly type experiment is to split the beam at a
$50:50$ beam splitter -- which introduces the vacuum mode $\vop v$ -- and then simultaneously
perform a homodyne detection at each output port. Measuring the amplitude
quadrature $\hat x_{\rm sim} = (\hat x + \hat v_1)/\sqrt{2}$ in one detector
and the phase quadrature $\hat p_{\rm sim} = (\hat p - \hat v_p)/\sqrt{2}$ in the other leads to
\begin{equation}
    \frac{\var{\hat x_{\rm sim}(t)}}{\absq{X_{\rm sim}(t)}}
    \frac{\var{\hat p_{\rm sim}(t)}}{\absq{P_{\rm sim}(t)}}
    \geq
    \frac{1+\cosh(2r)}{2\absq{X(t)}\absq{P(t)}} \,,
\end{equation}
where $X_{\rm sim}(t) = X(t)/\sqrt{2}$ and $P_{\rm sim}(t) = P(t)/\sqrt{2}$,
since also the signal is divided at the beam splitter.  Equation (4) states
that the achievable minimum uncertainty is indeed at least four times larger
than the limit imposed by Eq. (3). Squeezing does not help in this measurement
scenario and the best sensitivity is achieved with vacuum input, i.e.\ $r=0$.

For QDM, consider two squeezed vacuum modes with squeezing parameters $r_a$ and
$r_b$, described by the quadrature vectors $\vop a$ and $\vop b$ . For
simplicity, we restrict ourselves to the case where beam $\vop a$ is always
squeezed in the amplitude quadrature and the other beam is rotated in the
quadrature space by $\theta$. After entangling these beams at a $50:50$ beam
splitter, one mode is used as the meter state $\vop m$ which is modulated with
the signals $X(t)$, $P(t)$ as above, while the other state $\vop r$ is kept as
a reference beam. These two modes are then recombined at another beam splitter,
whose two output fields are sent to balanced homodyne detectors A and B
measuring the quadrature fields
\begin{equation}
    \hat x^{\rm ent}_{\rm sim} = \frac{1}{\sqrt{2}}(\hat r_1 - \hat x) = \hat
    a_1 - X^\text{ent}_\text{sim}(t) \,,
\end{equation}
and
\begin{equation}
    \hat x^{\rm ent}_{\theta,{\rm sim}} = \frac{1}{\sqrt{2}}(\hat r_\theta +
    \hat x_\theta) = \hat
    b_\theta + X^\text{ent}_{\theta,\text{sim}}(t) \,,
\end{equation}
respectively. Here, $\hat x_\theta = \hat x \cos\theta + \hat p \sin\theta$,
$X_\theta(t) = X(t)\cos\theta + P(t)\sin\theta$ and the signals are again
scaled by $1/\sqrt{2}$ since they are equally divided between the two homodyne
detectors. The corresponding uncertainty product reads
\begin{equation}
    \frac{\var{\hat x_{\rm sim}^{\rm ent}(t)}}{\absq{X(t)}/2}
    \frac{\var{\hat x_{\theta, {\rm sim}}^{\rm ent}(t)}}{\absq{X_\theta(t)}/2}
   = 
    \frac{e^{-2r_a}e^{-2r_b}}{\absq{X(t)}\absq{X_\theta(t)}} \,.
\end{equation}
This uncertainty is not bounded from below, for $r_a,r_b\to\infty$. Thus we
have shown that quantum-dense metrology, which is performed in relation to a
reference beam, can in principle reach arbitrarily high signal-to-noise
ratios.

\end{document}